\documentclass[aps,nofootinbib,preprintnumbers,prd,twocolumn]{revtex4-1}

\usepackage[T1]{fontenc}
\usepackage[utf8]{inputenc}

\usepackage{amsfonts}
\usepackage{amsmath}
\usepackage{graphicx}
\usepackage[svgnames]{xcolor}
\usepackage{hepparticles}
\usepackage{hepnicenames}
\usepackage{hepunits}
\usepackage{hyperref}
\usepackage{graphicx}
\usepackage{epstopdf}

\newcommand{\dual}[1]{\tilde{#1}}
\newcommand{\est}[1]{\widehat{#1}}
\newcommand{\ie}{\textit{i.e.}}
\newcommand{\nuvec}{\vec{\nu}}
\newcommand{\order}[1]{\mathcal{O}\left({#1}\right)}
\newcommand{\refapp}[1]{appendix~\ref{app:#1}}
\newcommand{\refeq}[1]{eq.~(\ref{eq:#1})}
\newcommand{\reffig}[1]{figure~\ref{fig:#1}}
\newcommand{\refsec}[1]{section~\ref{sec:#1}}

\newcommand{\rmdx}[1]{\mbox{d} #1 \,} 
\newcommand{\subd}{_{\text{d}}}
\newcommand{\subt}{_{\text{t}}}
\newcommand{\thvec}{\vec{\vartheta}}

\renewcommand{\theta}{\vartheta}
\let\eps\varepsilon
\newcommand{\vecest}[1]{\widehat{\vec{#1}}}

\newcommand{\xd}{x\subd}
\newcommand{\xt}{x\subt}
\newcommand{\Ekernel}{E(\xd|\xt)}

\DeclareMathOperator{\cov}{Cov}

\begin{document}

\allowdisplaybreaks

\preprint{SI-HEP-2014-17}
\title{Extracting Angular Observables without a Likelihood\\and Applications to Rare Decays}
\author{Frederik Beaujean}
\email{frederik.beaujean@lmu.de}
\affiliation{C2PAP, Universe Cluster, Ludwig-Maximilians-Universit\"at M\"unchen, Garching, Germany}
\author{Marcin Chrz\k{a}szcz}
\email{mchrzasz@cern.ch}
\author{Nicola Serra}
\email{nicola.serra@cern.ch}
\affiliation{Physik-Institut, Universit\"at Z\"urich, Z\"urich, Switzerland}
\author{Danny van Dyk}
\email{vandyk@tp1.physik.uni-siegen.de}
\affiliation{Theoretische Physik 1, Naturwissenschaftlich-Technische Fakult\"at,
Universit\"at Siegen, Siegen, Germany}

\begin{abstract}
  Our goal is to obtain a complete set of angular observables arising
  in a generic multi-body process. We show how this can be achieved
  without the need to carry out a likelihood fit of the angular
  distribution to the measured events. Instead, we apply the method of
  moments that relies both on the orthogonality of angular functions
  and the estimation of integrals by Monte Carlo techniques. The big
  advantage of this method is that the joint distribution of \emph{all}
  observables can be easily extracted, even for very few events. The
  method of moments is shown to be robust against mismodeling of the
  angular distribution. Our main result is an explicit algorithm that
  accounts for systematic uncertainties from detector-resolution and
  acceptance effects. Finally, we present the necessary
  process-dependent formulae needed for direct application of the
  method to several rare decays of interest.
\end{abstract}

\maketitle

\section{Introduction}
\label{sec:intro}

Our initial motivation for studying what we wish to call the
\emph{method of moments} is the determination of angular observables
in the rare FCNC-mediated decay $\bar{B}\to \bar{K}^*(\to
\bar{K}\pi)\ell^+\ell^-$. However, the method we describe in the
following is general and applies to all decay or scattering
processes that can be formulated in terms of an orthogonal basis of angular functions.
We find a previous work \cite{Dighe:1998vk} that
advocates this method chiefly for the determination of angular
observables in non-leptonic $B$ decays but also mentions the
applicability to semileptonic decays. Our aim is to improve upon this
previous work by studying the uncertainties that are introduced by
mismodeling of the angular distribution, and by working out a recipe
to determine and unfold detector effects. The latter is crucial for
the application of the method to real data. We show that
the method of moments has several major advantages over the usual approach
based on likelihood fits:
\begin{enumerate}
    \item Likelihood fits have convergence problems for a small number of
        events, and can require reparametrizations and/or approximations for a
        successful fit to the signal PDF. As an example, see the LHCb
        analysis of the angular distribution in $\bar{B}\to
        \bar{K}^*\mu^+\mu^-$ decays \cite{Aaij:2013qta}.

        The method of
        moments does not require any such reparametrizations or approximations.

    \item Likelihood fits  can be unstable in case the underlying
        physical model is only partially known. This can lead to
        overestimating the number of physical parameters, and consequently
        inhibits the convergence of the fits. As an example, this type
        of problem occured in toy studies of the decay $B\to K^*(\to K\pi)\ell^+\ell^-$
        as reported in \cite{Egede:1142152}. It was subsequently solved
        when a missing symmetry relation between the angular observables
        was found and applied, thereby reducing the number of fit parameters.

        In contrast, we will show that the method of moments does not require
        information on the correlations between model parameters as an input.
        Instead, it yields the correlations as an output, at the expense of
        somewhat larger uncertainties.

    \item Mismodeling the underlying physics model can result in systematic bias
        in likelihood fits.

        We will
        show that the method of moments is insensitive to a certain type of mismodeling;
        i.e. introducing a cutoff in a partial-wave expansion of the signal PDF.

    \item Using the method of moments, the joint probability distribution of the
        angular observables rapidly converges towards a multivariate Gaussian distribution.
        This allows an easy transfer of correlation information from the experiments to
        interested theorists.
\end{enumerate}

We continue with basic definitions that pertain to angular observables, and our
results in the subsequent sections. Let $\thvec$ denote the set of all
angles, and let $\nuvec$ denote the set of all
other non-angular kinematic variables needed to fully specify the
final state of the process under study. For example, $\nuvec$ may
include invariant masses or center-of-mass energies. We define an
angular observable $S_i$ as a coefficient in the probability density
function (PDF), $P(\nuvec, \thvec)$, of the process by
\begin{align}
    \label{eq:def-P}
    P(\nuvec, \thvec) \equiv \sum_i S_i(\nuvec) \times f_i(\thvec)\,.
\end{align}
Here, the dependence on the decay angles $\thvec$ has been
explicitly factored out in terms of the angular functions
$\{f_i(\thvec)\}$. We assume there exists a dual basis of functions
$\{\dual{f}_i(\thvec)\}$ such that the orthonormality relations
\begin{equation}
    \label{eq:def-ortho-rel}
    \int_\Omega \rmdx{\vec{\theta}} \dual{f}_i(\thvec) f_j(\thvec)  = \delta_{ij}
\end{equation}
hold with $\Omega$ representing the full angular phase space relevant
to the process.  For particle decays, $P$ is generally expressed in
terms of the fully differential decay width,
\begin{align}
    \label{eq:def-P-decay}
    P(\nuvec, \thvec) \equiv \frac{1}{\Gamma}\frac{\rmdx{^{2}\Gamma}}{\rmdx{\vec\nu} \rmdx{\vec\theta}}\,,
\end{align}
where $\Gamma$ is the total decay width. For a scattering process, one can similarly use
\begin{align}
    \label{eq:def-P-scattering}
    P(\nuvec, \thvec) \equiv \frac{1}{\sigma}\frac{\rmdx{^{2}\sigma}}{\rmdx{\vec\nu} \rmdx{\vec\theta}}\,,
\end{align}
where the total cross section $\sigma$ is used for the
normalization. Since the determination of the total decay width or
total cross section can be quite difficult, we emphasize that
different normalizations for $P$ can be used.  For instance, the total
decay width (or cross section) of the process of interest can be
replaced by the corresponding quantity of a control-channel
process. This change of normalization is equivalent to a linear rescaling of
the angular observables $\lbrace S_i\rbrace$; thus ratios or similar suitable combinations
of the angular observables are not affected by a change of normalization.\\

Our method is an extension of the classical method of moments with
orthogonal functions~\cite[sec. 8.2]{James:2006zz}. The only difference
is that conventionally the angular functions are assumed
\emph{self-dual}, $\dual{f}_i = f_i$. However, it suffices that the
system of angular functions $\{f_i(\thvec)\}$ \emph{can} be transformed into
an orthonormal basis. We find it convenient to work
in the basis of Legendre polynomials that are not self-dual.
Our approach covers the self-dual case, provided that one
replace $\dual{f}_i \to f_i$ appropriately.
Using the ansatz
\begin{equation}
  \label{eq:dual-ansatz}
  \dual{f}_i = \sum_{j} a_{ij} f_j \,,
\end{equation}
the dual basis needs to be worked out case by case through solving the
linear system of equations (\ref{eq:def-ortho-rel}). For a selection
of hadron decays with a $b$ quark in the initial state and two leptons
in the final state, we list the dual bases in a series of appendices
\ref{app:btokll} through \ref{app:btokstarll}. Note that a
similar analysis was done in~\cite{Dighe:1998vk} for the decays $B \to
J/\psi \phi$ and $B \to J/\psi K^{*}$.

In the remainder of this letter we discuss how to obtain an angular
observable $S_i(\nuvec)$ in an experimental setup where each recorded
event is (approximately) distributed according to $P$.  We establish
the statistical basics in section \ref{sec:sample-based-det}. Section
\ref{sec:systematics} is dedicated to the impact of systematic effects
such as mismodeling the underlying physics or detector acceptance
effects. Numerical studies for one uni-angular and one triple-angular
distribution are provided in section \ref{sec:numerics}.

\section{Sample-Based Determination}
\label{sec:sample-based-det}

The orthonormality relations \refeq{def-ortho-rel} imply that a single angular observable $S_i$
can be projected out of the full PDF $P$ as
\begin{equation}
    \label{eq:det-Pi-analytical}
    S_i(\nuvec) = \int_{\Omega} \rmdx{\vec\theta}  P(\nuvec, \thvec) \dual{f}_i(\thvec)\,.
\end{equation}
where $\lbrace \dual{f}_i \rbrace$ denotes a \emph{dual basis} of angular
functions, and $\Omega$ represents the entire angular phase space. In general,
$\lbrace \dual{f}_i \rbrace$ may differ from $\lbrace f_i \rbrace$. This is the
case for our selection of applications in appendices \ref{app:btokll} through
\ref{app:btokstarll}.\\

It is sensible to refer to the angular observable $S_i$ as the
\emph{$f_i$-moment} of the PDF $P$.  We emphasize that a relation of
type \refeq{det-Pi-analytical} holds for any combination of a density
written as in \refeq{def-P} and an orthonormal basis of angular
functions $\lbrace f_i \rbrace$; \ie, there is no unique
basis of angular functions. For the proof we refer to ref. \cite{Dighe:1998vk}.\\

Integration over the non-angular variables yields
\begin{equation}
    \langle S_i\rangle
    \equiv \int \rmdx{\vec\nu} S_i(\nuvec)
    = \int \rmdx{\vec \nu} \left[\int_{\Omega} \rmdx{\vec \theta} P(\nuvec,\thvec) \dual{f}_i(\thvec) \right].
\end{equation}
The remainder of this section describes the method of moments, in
which we replace the analytical integration by Monte Carlo (MC)
estimates.  The central tenet of MC integration is the fact that the
expectation value $E_P[g]$ of some function $g(x)$ under the
probability density $P(x)$,
\begin{equation}
    E_P[g] \equiv \int \rmdx{x} P(x) g(x),
\end{equation}
can be approximated by the consistent and unbiased
estimator $\est{E_P[g]}$~\cite[sec. 8.2]{James:2006zz}
\begin{equation}
    \label{eq:mc-id}
    E_P[g] \to \widehat{E_P[g]} \equiv \frac{1}{N} \sum_{n=1}^N g(x^{(n)}) \,,\,    x^{(n)} \sim P
\end{equation}
due to the strong law of large numbers for $N \to \infty$, assuming
that the variates $x^{(n)}$, $n = 1, \dots, N$, are distributed as
$P$.
Throughout this letter we denote all MC estimators with a wide hat.\\

Application of \refeq{mc-id} then yields
\begin{equation}
    \langle S_i\rangle \to \widehat{\langle S_i\rangle} = \frac{1}{N} \sum_{n=1}^{N} \dual{f}_i(x^{(n)})\,.
\end{equation}
It is often of interest to obtain observables integrated over certain
bins of $\nuvec$. We define
\begin{align}
    \langle S_i\rangle_{\vec{a},\vec{b}}
    & \equiv \int_{\vec{a}}^{\vec{b}} \rmdx{\nuvec} S_i(\nuvec)\\
    & = \int_{\vec{a}}^{\vec{b}} \rmdx{\nuvec} \left[\int_{\Omega} \rmdx{\vec\theta} P(\nuvec,\thvec) \dual{f}_i(\thvec)  \right]\\
    & = \int \rmdx{\nuvec} \left[\int_{\Omega} \rmdx{\vec\theta}  P(\nuvec,\thvec) \dual{f}_i(\thvec)
        \mathbf{1}(\vec{a} \le \nuvec \le \vec{b})\,
        \right],
\end{align}
where the argument of the indicator function $ \mathbf{1}(\vec{a} \le
\nuvec \le \vec{b})$ is to be interpreted componentwise.
Application of \refeq{mc-id} immediately yields
\begin{equation}
    \label{eq:bin-importance}
    \widehat{\langle S_i\rangle_{\vec{a},\vec{b}}}
    = \frac{1}{N} \sum_{n=1}^{N} \dual{f}_i(x^{(n)})         \mathbf{1}(\vec{a} \le \nuvec \le \vec{b})\,.
\end{equation}
For notational simplicity, let us forget about the $\nuvec$
integration and consider only $S_i$. In the limit $N \to \infty$, the
central limit theorem (CLT) implies that the random vector
\begin{equation}
  \label{eq:angular-obs-vec}
  \vecest{S} \equiv (\est{S_0}, \dots, \est{S_i}\,,
  \dots)
\end{equation}
follows a multivariate Gaussian distribution $\mathcal{N}(\vec{S},
\Sigma)$ centered on the true value $\vec{S}$ with the
covariance $\Sigma_{ij}$ estimated as
\begin{equation}
\begin{aligned}
    \Sigma_{ij}
        & \equiv \cov[S_i,S_j]\\
        & \to \est{\Sigma_{ij}} \equiv \est{\cov}[{S}_i, {S}_j]\\
        & = \frac{1}{N - 1} \sum_{n=1}^{N} \Big[\dual{f}_i\big(x^{(n)}\big) - \est{S_i}\Big]\,\Big[\dual{f}_j\big(x^{(n)}\big) - \widehat{S_j}\Big]\,.
\end{aligned}
\end{equation}
In our physics applications, the parameter space is compact and each
$\dual{f}_i$ is bounded. Hence the requisites for the most basic
version of the CLT to hold --- finite mean and covariance of
$\dual{f}_i$ --- are automatically satisfied.
In our numerical analysis the sample covariance rapidly converges towards the true covariance
matrix; see also \refsec{numerics}.\\

Compared to the usual maximum-likelihood approach, we find for the method of moments:
\begin{enumerate}
  \item The angular observable ${S_i}$ can be determined
  independently of any other observable ${S_j}$. It is therefore
  much more robust to physics assumptions needed to define the full
  likelihood. In particular, this means one does not have to be
  specific regarding the form of new-physics contributions; in fact,
  one does not even need to be able to explicitly formulate the
  likelihood at all.
\item It is superior for a small number of samples $N$. Likelihood
  fits tend to be numerically unstable if lots of parameters need to
  be estimated from sparse data. This is more severe if the mode of
  the likelihood is near the boundary of the physically allowed
  region \cite{lehmann1998}. For some of these decays of interest, there are only $\order{100}$ events
  recorded per bin.

\item The estimate is unbiased for any $N$. In contrast, the
  maximum-likelihood estimate has a bias of order
  $1/N$~\cite{Cox:1968}. In practice, one should keep in mind the
  bias-variance trade-off: it is a well known phenomenon that removing
  the bias usually leads to an increase in variance of the sampling
  distribution of the estimator and vice versa~\cite[sec. 7.3]{James:2006zz}. From a
  Bayesian decision-theory point of view, both contribute similarly to
  the expected loss associated with deciding on just one value of the
  unknown parameter. One should therefore not prefer the method of moments
  over likelihood fits just because the former reduces
  the bias~\cite[sections 13.8,17.2]{jaynes:2003}. In fact, for
  the results discussed below in \refsec{numerics}, the likelihood
  fits --- if they converge --- exhibit a negligible bias and
  produce uncertainties $10\%$--$30\%$ smaller than those from the method of moments.

\item The approximate multivariate Gaussian distribution of $\vecest{S}$
    allows easier and more precise transfer of the information in the
  data to interested theorists for more accurate fits of
  standard-model and new-physics parameters \cite{Altmannshofer:2013foa,
Descotes-Genon:2013wba,Beaujean:2013soa},
  or for more precise predictions of optimized observables; see e.g.,
  \cite{Egede:2008uy,Egede:2010zc,Bobeth:2010wg,Becirevic:2011bp,
    Bobeth:2012vn,Matias:2012xw,DescotesGenon:2012zf} for definitions
  of such optimized observables in $B\to K^*\ell^+\ell^-$ decays,
  \cite{Faller:2013dwa} for application to the decay $B\to
  \pi\pi\ell^-\bar\nu_\ell$, and \cite{Boer:2014kda} for observables
  in $\Lambda_b\to\Lambda(\to N\pi)\ell^+\ell^-$). While the
  likelihood also approaches a multivariate Gaussian as $N \to
  \infty$, the two methods differ in their utility as input for
  theorists if $\widehat{\vec{S}}$ is not well
  inside the physical region. For example, suppose there are two
  angular observables that are constrained to a triangular region by
  phase-space or unitarity arguments as
  \begin{equation}
    \label{eq:constr-ex}
    |S_1| + S_2 \le 1, \, S_1 \in [-1,1],\, S_2 \in [0,1] \,.
  \end{equation}
  It may (and often does) happen in practice that $\widehat{\vec{S}}$
  is close or even outside the allowed region such that a significant
  part of the probability mass covers unphysical values. In a Bayesian
  fit, one would take $\mathcal{N}(\vecest{S} | \vec{S},
  \widehat{\Sigma})$ as the sampling distribution of the ``data'', and
  simply set a uniform prior on the triangle in the $\vec{S}$ plane
  defined by \refeq{constr-ex} to have a well defined problem. This
  could be trivially combined with other independent information in a
  global fit. Someone with a different physics model might have to
  consider a
  different physical region, and could incorporate it just as easily.\\

  For a likelihood fit, the constraint needs to be part of the analysis
  performed by the experimental collaboration, and the resulting likelihood as
  a function of $\vec{S}$ may be distinctly not Gaussian.  Communicating such a
  result has proved to be challenging due to technical reasons (such as data
  formats, size etc.) This leads to the undesirable situation that only the
  mode and standard errors are reported, and theorists often include the
  results as independent measurements with a Gaussian distribution and
  disregard the boundary problem as well as correlations altogether.
\end{enumerate}

\section{Sources of Systematic Uncertainties}
\label{sec:systematics}

In \refsec{sample-based-det}, we assume that the PDF $P$ describes the underlying physics accurately,
and that the experiment observes each event with perfect accuracy. In order to estimate systematic
uncertainties, we lift these assumptions.

\subsection{Mismodeling due to Contributions by Higher Partial Waves}
\label{sec:systematics:partial-waves}

\begin{figure}
    \includegraphics[width=.45\textwidth]{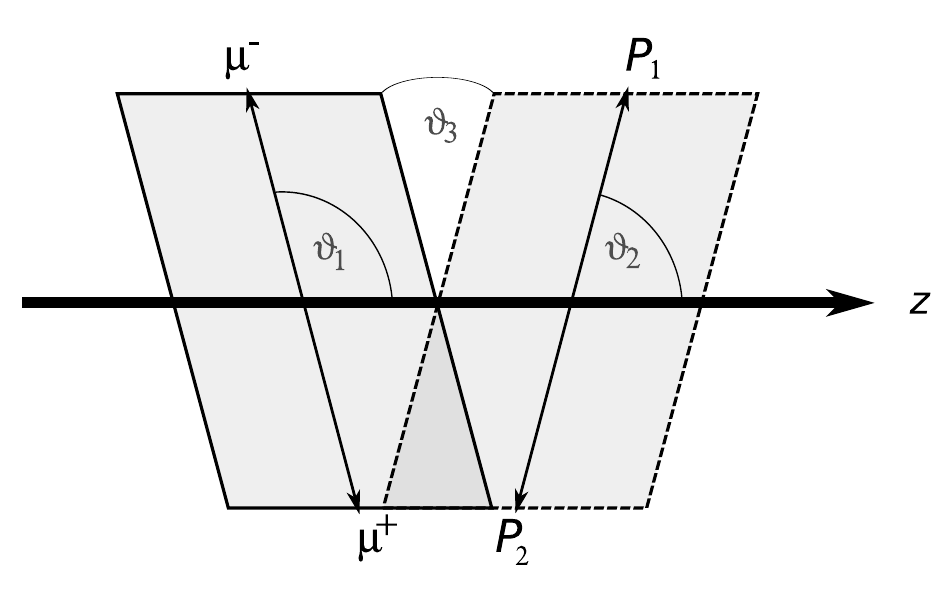}
    \caption{Decay topology for decays $B\to P_1 P_2 \ell_1 \bar\ell_2$. \label{fig:topology}}
\end{figure}

In several interesting processes we might only have an approximate
result for $P$.  In this section, we focus on one particular class of
mismodeling of the signal PDF: the angular-momentum cutoff in partial-wave expansions. This mismodeling potentially affects a large number
of decays and scattering processes. For the sake of clarity we take
the interesting class of four-body decays $B\to P_1 P_2 \ell_1 \ell_2$
as an example\footnote{This includes the rare $b\to s$ mediated $B$ decay $B
  \to K\pi\ell^+\ell^-$, and the $V_{ub}$ suppressed decay $B\to
  \pi\pi\ell^+\bar\nu_\ell$. For both examples the PDF $P$ is known in
  the small-width approximation and when assuming a pure $P$-wave
  resonant final state. An extension to $S-P$ interference has been
  studied for $B\to K\pi\ell^+\ell^-$
  \cite{Blake:2012mb,Becirevic:2011bp}, and for $\bar{B}\to
  \pi\pi\ell^-\bar\nu_\ell$ \cite{Faller:2013dwa}. For a first study
  of $S$, $P$
  and $D$ interference, see \cite{Das:2014sra}
}.\\

Within existing analyses, the PDFs of these decays are usually expressed in terms of one or a few
partial waves of the dimeson system. However, the angular momentum of the dimeson system is unbounded
from above, and gives rise to an infinite set of angular observables.\\

For the selected class of decays, we can describe that problem as follows: The
PDF $P$ has a fixed dependence on the dilepton helicity angle $\theta_1$ and
the azimuthal angle $\theta_3$. (See also \refapp{btokstarll} for details on
the angular distribution.) However, at the level of decay amplitudes the
dimeson system can have an arbitrarily large total angular momentum $j$; only
its third component is restricted to $j_z = -1,0,+1$.  It is then convenient to
compute the angular distribution explicitly in terms of $\theta_1$ and
$\theta_3$, but leave the angular observables dependent on the remaining
helicity angle $\theta_2$ of the dimeson:
\begin{multline}
    P(\cos\theta_1,\cos\theta_2,\theta_3)\\
        \equiv \sum_k S_k(\nuvec,\cos\theta_2) f_k(\cos\theta_1,\theta_3)\,.
\end{multline}
We advocate here that is a sensible procedure to
perform an expansion in terms of Legendre polynomials $p_{l}^{(|m|)}$ with
respect to the remaining angle $\theta_2$. Here, the angular-momentum indices
$l$ and $m$ follow from the usual rules for addition of the angular momenta $j$
and $\tilde{j}$ of the partial-wave expansion of the underlying amplitude and
its complex conjugate: $j - \tilde{j} \leq l \leq j + \tilde{j}$, and $m = j_z
+ \tilde{j}_z$.

For the decay at hand, we consider the partial-wave expansion for the angular observables (see \cite{Lee:1992ih} and \refapp{partial-waves})
\begin{equation}
    S_k(\vec{\nu}) \equiv \frac{I_k(\nuvec)}{4\pi \Gamma}\,,\quad\text{with}\quad \Gamma = I_1(\nuvec) - \frac{I_2(\nuvec)}{3}\,,
\end{equation}
which reads
\begin{multline}
    \label{eq:expansion}
    S_{k}(\vec{\nu},\cos\theta_2) \\
        \equiv \sum_{l=0}^\infty \frac{1}{n_{l,|m|}} S_{k,l}(\vec{\nu}) p_{l}^{(|m|)}(\cos\theta_2)\,.
\end{multline}
The normalization factor $n_{l,|m|}$ is defined in \refeq{legendre-scalar-product}. Within our
example we have (cf. also \refapp{btokstarll} and \cite{Lee:1992ih})
\begin{equation}
    |m| = \begin{cases}
        0\,, & k = 1,2,6\\
        1\,, & k = 4,5,7,8\\
        2\,, & k = 3,9
    \end{cases}\,.
\end{equation}

The angular observables $S_{k,l}$ -- as defined in \refeq{expansion} -- have
the merit of a well defined total angular momentum, and thus are physically
distinguishable.  As a consequence of the orthogonality of the Legendre
polynomials, any mismodeling (or rather, lack of modelling) of higher
partial-wave observables \emph{does not affect} the method of moments as
discussed in the previous section. That is to say, adding further (orthogonal) terms
to the PDF only appends observables to $\vec{S}$, but does not change the leading
elements. The same applies to the covariance.\\

Unfortunately, this benefit on the experimental side is accompanied with a
theoretical draw back. Each observable $S_{k,j}(\vec{\nu})$ consists of an
infinite sum of bilinears of partial-wave amplitudes. It remains for
theoretical analyses to estimate or calculate the impact of partial waves
beyond the S and P wave contributions.  (For $B\to K\pi\ell^+\ell^-$ a first
study has been carried out where contributions up to the D wave are
investigated \cite{Das:2014sra}).\\

We wish to emphasize that detector acceptance effects systematically affect the
expansion for any basis of angular functions, including the one suggested in
this section.  Nevertheless, the expansion in terms of Legendre polynomials as
suggested above provides means to cope with these effects, as we discuss in the
following subsection.

\subsection{Recipe for Including Detector Effects}
\label{sec:systematics:acceptance}

Ascertaining a detector's performance to detect signal events
with accurate determination of the event's angles is generally a
difficult task. In the ideal case, one would have an explicit
probabilistic model of the detector acceptance and could thus
write down the full forward model from which the measured events
arise. In practice, that is not feasible, and one is
forced to simplify the model.
The standard approach is to generate the \emph{true} particle events
$\lbrace x_\text{t}^{(n)}\rbrace =
\lbrace(\vec{\nu}^{(n)}_\text{t},\vec\theta^{(n)}_\text{t})\rbrace$,
$n=1,\dots,N_\text{t}$, from a PDF assumed to describe the bare
physical process, and to propagate those particles through a detailed
simulation of the detector.  The observable traces that the particles
leave in the detector are fed into reconstruction algorithms resulting
in the \emph{detector} events $\lbrace x^{(n)}_\text{d}\rbrace =
\lbrace(\vec{\nu}^{(n)}_\text{d}, \vec\theta^{(n)}_\text{d})\rbrace$,
$n=1,\dots,N_\text{d}$ with $N_\text{d} \le N_\text{t}$.  In general,
the distribution of the detected events is
\begin{equation}
    P_\text{d}(x_\text{d}) = \frac{1}{R} \int \rmdx{x_\text{t}} P_\text{t}(x_\text{t}) E(x_\text{d} | x_\text{t}) \,.
\end{equation}
Here $P_\text{t}$ is the probability distribution of the true events,
the normalization constant $R$ is given by
\begin{equation}
  \label{eq:def-R}
    R \equiv \iint \rmdx{x_\text{t}} \rmdx{x_\text{d}} P_\text{t}(x_\text{t}) E(x_\text{d} | x_\text{t})\,.
\end{equation}
The kernel $E(x_\text{d}| x_\text{t})$ is usually decomposed as
\begin{equation}
  \Ekernel = \eps(\xt) P(\xd|\xt),
\end{equation}
where the PDF $P(\xd|\xt)$ models the resolution effects and the
unnormalized density $\eps(\xt)$ is the \emph{detector acceptance}
function. Perfect resolution corresponds to
\begin{equation}
  P(\xd|\xt) = \delta(\xd - \xt) \,.
\end{equation}

In what follows, we propose a systematic method to unfold all effects
of $\Ekernel$ through MC simulations and the method of moments,
using that $\Ekernel$ can be expanded --- at
least formally --- in Legendre polymials.  For illustration, we
proceed with the explicit example of a uniangular\footnote{%
  The generalization of this section to multiangular PDFs is
  straighforward. It can be achieved by promoting $x$ to a vector,
  promoting the Legendre polynomials to products of independent
  polynomials or spherical harmonics, and promoting the indices
  $i,j,k,m$ to multi-indices.  } PDF with $x = \cos\theta$. Let us
define the PDF in terms of the Legendre polymials (i.e., $f_k(x)
\equiv p_k(x)$) and angular observables $\vec{S}$ as
\begin{equation}
    P\subt(\xt) \equiv P\subt(\xt | \vec{S}) = \sum_k S_k f_k(\xt),
  \end{equation}
where $k = 0, 1, \dots$ denotes an angular-momentum-like index associated with the observables.
Normalization of $P\subt$ is equivalent to choosing $S_0 = 1/2$. Requiring $P\subt(\xt) \ge 0 \,\forall \xt$
implies $|S_k| \leq 1/2$ for $k > 0$. More stringent relations between the
$S_k$ might hold, but are of no concern here. For later use, we define
\begin{equation}
  \label{eq:def-Sk}
  S^{(m)}_k \equiv
  \begin{cases}
    1/2 \delta_{k,0}, & m=0\\
    1/2 (\delta_{k,0} + \delta_{k,m}), & m>0 \,,
  \end{cases}
\end{equation}
and note that
\begin{equation}
    P^{(m)}(\xt) \equiv P(\xt | \vec{S}^{(m)})
\end{equation}
is a valid PDF. The dual basis of angular functions follows then from the normalization
and orthogonality of the Legendre polynomials, and one therefore has
\begin{gather}
    \tilde{f}_k(x) = \frac{2 k + 1}{2} f_k(x)\,,\\
    \int_{-1}^{+1} \rmdx{x} \tilde{f}_k(x) f_l(x) = \delta_{k,l}\,.
\end{gather}
We now define the \emph{simulated raw} moments $\vec{Q}^{(m)}$ as
\begin{equation}
    \label{eq:analytic-raw-moments}
    Q_i^{(m)} \equiv \iint \rmdx{\xt} \rmdx{\xd} \tilde{f}_i(\xd) P^{(m)}(\xt) \Ekernel,
\end{equation}
which are instrumental to our recipe. Monte Carlo estimators of these moments can be constructed from specifically crafted detector events $x_\text{d}^{(n,m)}$, $n = 1, \dots, N_\text{d}^{(m)}$, where
\begin{equation}
    x_\text{d}^{(n,m)} \sim P_E^{(m)}(\xd) \equiv \frac{1}{R^{(m)}} \int \rmdx{\xt} P^{(m)}(\xt) \Ekernel\,.
\end{equation}
In words, for each $m$ it is required to generate events from
a toy physical distribution $P^{(m)}$, for which $S_0 = S_m=1/2$ and all other
observables are set to zero. Next, propagate these events through
a detector simulation. The normalization $R^{(m)}$ is chosen such
that $\int P_E^{(m)}(\xd) \rmdx{\xd} = 1$.  We emphasize that
$R^{(m)}$ can be estimated as $\est{R}^{(m)} = N_\text{d}^{(m)} /
N_\text{t}$, where $N_\text{t}$ corresponds to the number of
simulated true events.  The estimators then read
\begin{equation}
    \est{Q}_i^{(m)} \equiv \est{R}^{(m)} \frac{1}{N_\text{d}} \sum_n^{N_\text{d}} \tilde{f}_i(x_\text{d}^{(n,m)})
\end{equation}

Linearity of the integral over $x$ and convergence of the expansion of
$P_E^{(m)}$ in terms of Legendre polynomials ensures that
\begin{equation}
    \vec{Q}^{(m)} = M \vec{S}^{(m)}\,.
\end{equation}
We call the matrix $M^{-1}$ the \emph{unfolding matrix},
which is specific to the decay at hand.
Given our definition of $S_k^{(m)}$ in
\refeq{def-Sk} it is easy to see that
\begin{equation}
    M_{ij} = \begin{cases}
        2 Q_i^{(0)}                          & j = 0\,,\\
        2\left(Q_i^{(j)} - Q_i^{(0)}\right)  & j \neq 0\,,\\
    \end{cases}
\end{equation}
and its MC estimator $\est{M}$  can be obtained through the replacements $Q_i^{(m)} \to \est{Q}_i^{(m)}$.\\

In order to finally extract the angular observables from data, we use the \emph{measured raw moments}. Their MC estimator
$\vecest{Q}$ --- based on the \emph{detected events} $x^{(n)}$, $n=1,\dots,N$ --- reads
\begin{equation}
    \est{Q}_i \equiv R \frac{1}{N} \sum_n^N \tilde{f}_i(x^{(n)})\,.
\end{equation}
We then obtain MC estimators of the angular observables via
\begin{equation}
    \label{eq:estS}
    \vecest{S} \equiv \left[\est{M}^{-1}\right] \vecest{Q}\,.
\end{equation}

Apparently we now face a circular dependence. On the one hand,
the estimators $\vecest{Q}$ and thus also $\vecest{S}$
are proportional to $R$, the ratio of detected events over occuring events.
On the other hand, $R$ depends by construction (see \refeq{def-R})
on $P_\text{t}(x_t)$, and thus on the true value of the angular observables $\vec{S}$.
This dependence is broken by the fact that the MC estimators $\vecest{S}$ need to fulfill
the self-consistency condition
\begin{equation}
    \est{S}_0 = \frac{1}{2}\,,\qquad\text{for a uniangular distribution}.
\end{equation}
(For the process-dependent conditions in the multiangular case see \refeq{recipe:lambdabtolambdall} and
\refeq{recipe:btokstarll}.)
This self-consistency condition is tightly related to the determination of
the branching ratio of the underlying decay, and we therefore suggest to
carry out a combined analysis for the determination of the branching ratio
and the extraction of the angular observables.
Moreover, in the applications to $B$ decays only ratios and similar $R$-independent combinations
of the angular observables are of interest.
\\

We note that $\vecest{S}$ as determined from \refeq{estS} does not fully
correspond to $\vec{S}$ for arbitrary detector acceptance $\eps$.
Assuming an expansion of $\eps$ in terms of Le\-gendre polynomials up to a given
order $L$, we have to calculate the raw moments up to $\dim \est{Q} = \dim
\est{S} = \dim S + L$. The MC estimators of the corrected angular observables
then take the following structure:
\begin{equation}
    \vecest{S} = ( \underbrace{\est{S}_0, \dots, \est{S}_{\dim S - 1}}_\text{physical}, \underbrace{\est{S}_{\dim S}, \dots, \est{S}_{\dim Q - 1}}_\text{``superfluous''})\,.
\end{equation}
The method is consistent as long as the MC estimators for the ``superfluous''
observables are compatible with zero. \footnote{This holds only for the
    physical model of uniangular decay distributions. If the physical model
    involves a partial-wave expansion with cutoff, these superfluous
    observables correspond to higher partial waves that have been suppressed in
the physical model; see \refapp{btokstarll} for such a case.} The value $L$ depends
on the setup of the particle detector under consideration, and remains to
be determined just as in studies that carry out a likelihood fit.\\

The accuracy of the unfolding process as outlined above critically depends on
both the accuracy of the detector simulation, as well as the uncertainties
induced by the MC estimates.  For an experimental analysis, one would now turn
to the determination of the distribution of $\vecest{S}$ as a function of the
detector setup. This would involve the determination of both $M$ and
$\vecest{Q}$ for a number of detector configurations, and subsequent profiling
or marginalization.  While such considerations of any detector simulation are
beyond the scope of this work, we can, however, comment on the MC-induced
uncertainties. As usual, one needs to find a balance between compute time and
accuracy. For a uniangular distribution and $\order{10^6}$ MC samples, we find
that the error on the mean of each matrix element is $\order{10^{-4}}$.
This suggests that the so-induced systematic error can be driven below any
statistical uncertainty.\\
An alternative method to unfold detector effects is weighting the data on an
event-by-event basis, with each weight corresponding to the inverse of the detection
efficiency.\\

Let us conclude this section by commenting that parts of the unfolding matrix
are universal in a sense: They can be reused in analyses with a
similar underlying decay.  Therefore, computing resources spent on improving
the accuracy of $\est{M}$ are not wasted.

\section{Toy Studies}
\label{sec:numerics}

We now study the performance of the proposed method. In order to do so, we simulate
individual events for two separate physical processes: one uni-angular, and one tri-angular decay
distribution. We repeat the analysis for varying sample sizes, ranging from
$50$ to $500$ events. Our toy analyses are based on SM predictions for angular observables
in the decays $B\to K\ell^+\ell^-$ and $B\to K^*(\to K\pi)\ell^+\ell^-$.
In order to faithfully investigate the performance of the method of moments, we repeat our
numerical studies for several bins in the kinematic range $1\,\GeV^2 \leq q^2 \leq 6\,\GeV^2$,
as well as $15\,\GeV^2 \leq q^2 \leq q^2_\text{max}$. Here, the bin width is chosen either
as $1\,\GeV^2$ or $0.5\,\GeV^2$. This setup is meant to ensure that a wide spectrum of possible
values for the angular observables is investigated.\\

\begin{figure}[b]
        \centering
            \includegraphics[width=0.45\textwidth]{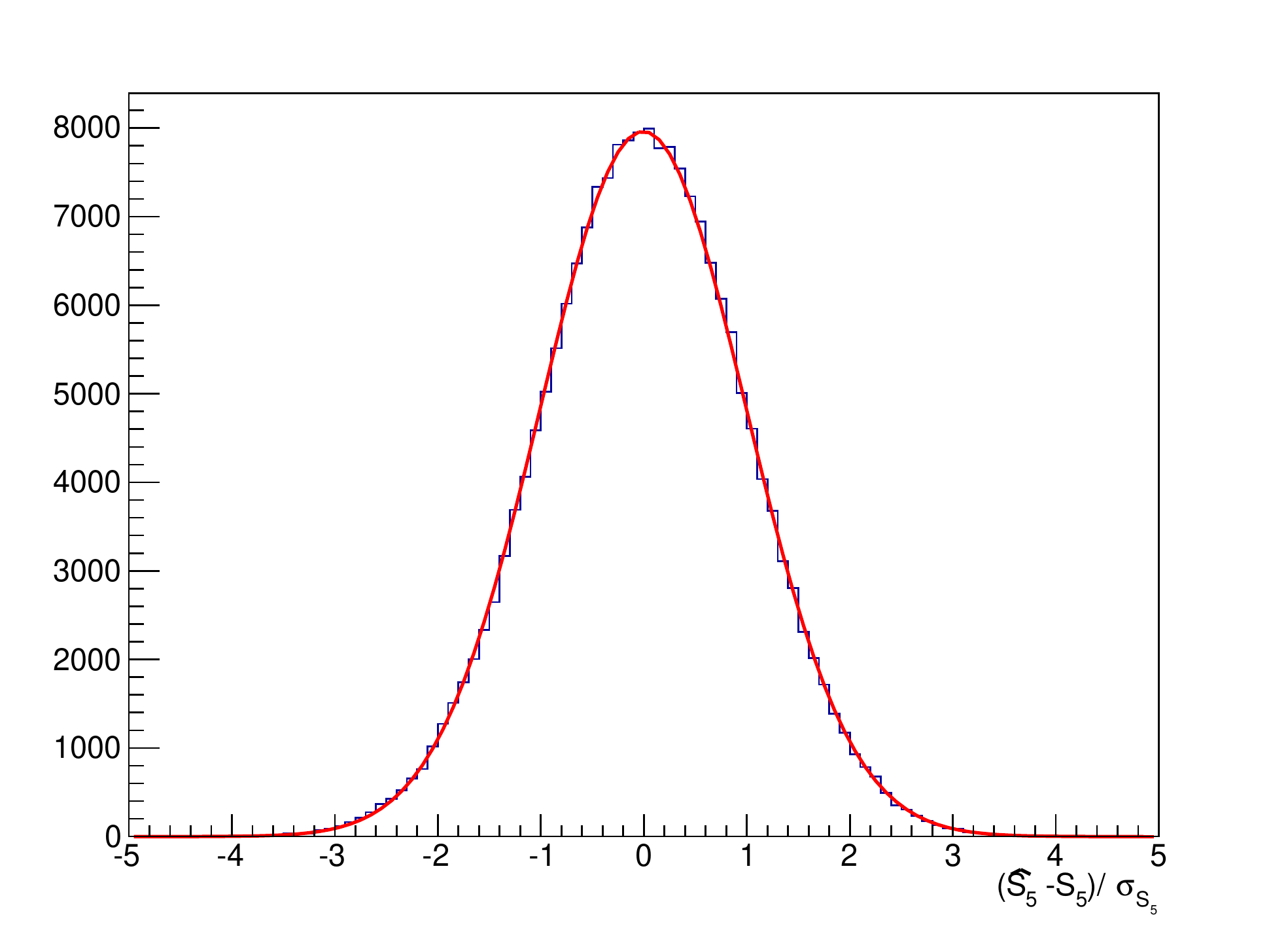}
            \includegraphics[width=0.45\textwidth]{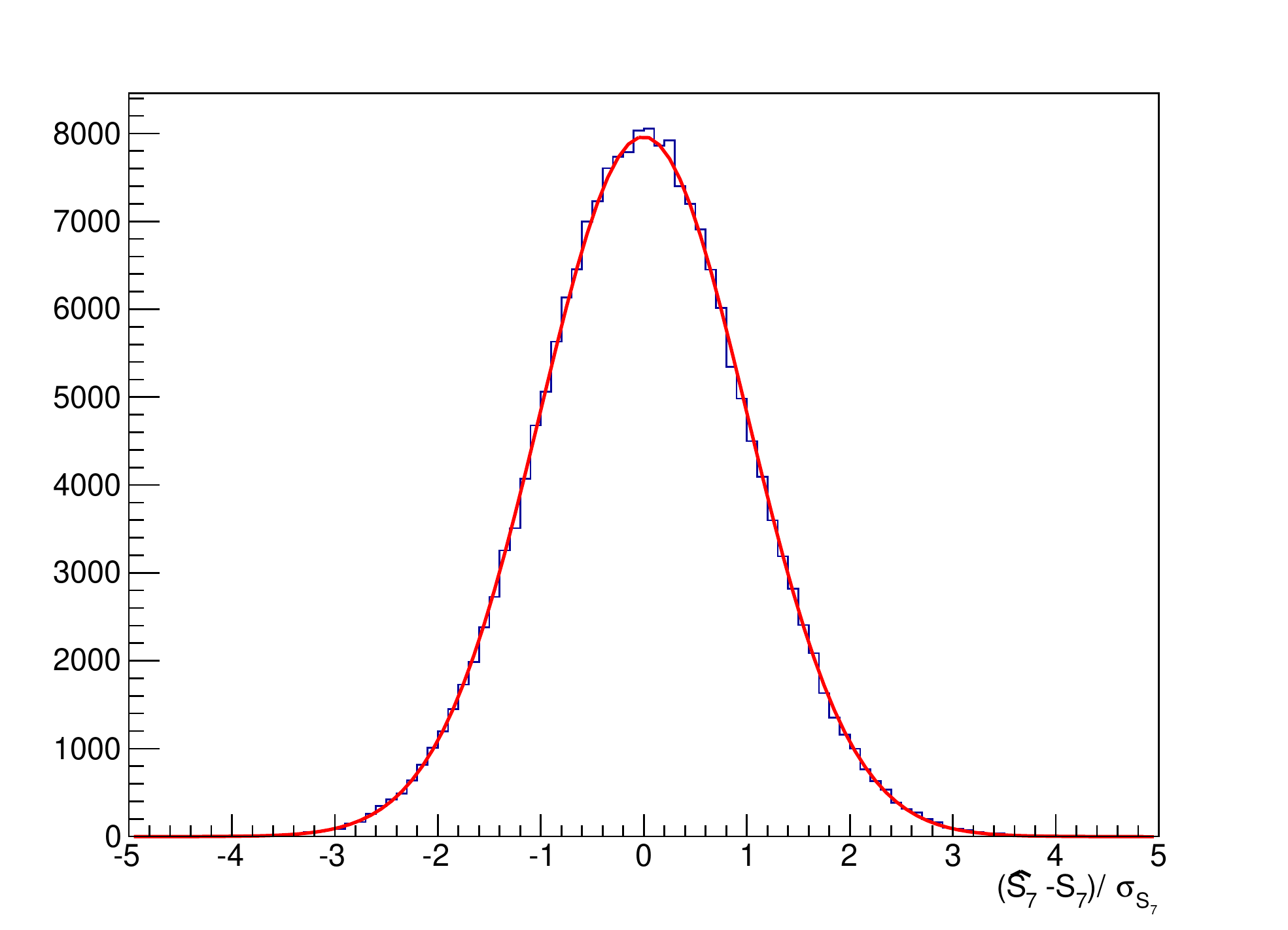}
        \caption{Pull distribution for the angular observables $S_5$ (upper) and $S_7$ (lower), extracted from $2\cdot 10^5$ studies of
        $200$ simulated events each for the decay $B\to K^*\ell^+\ell^-$. The red curve represents a fit to a Gaussian distribution.}
        \label{fig:pulls}
\end{figure}

\begin{figure}[t]
        \centering
            \includegraphics[width=0.45\textwidth]{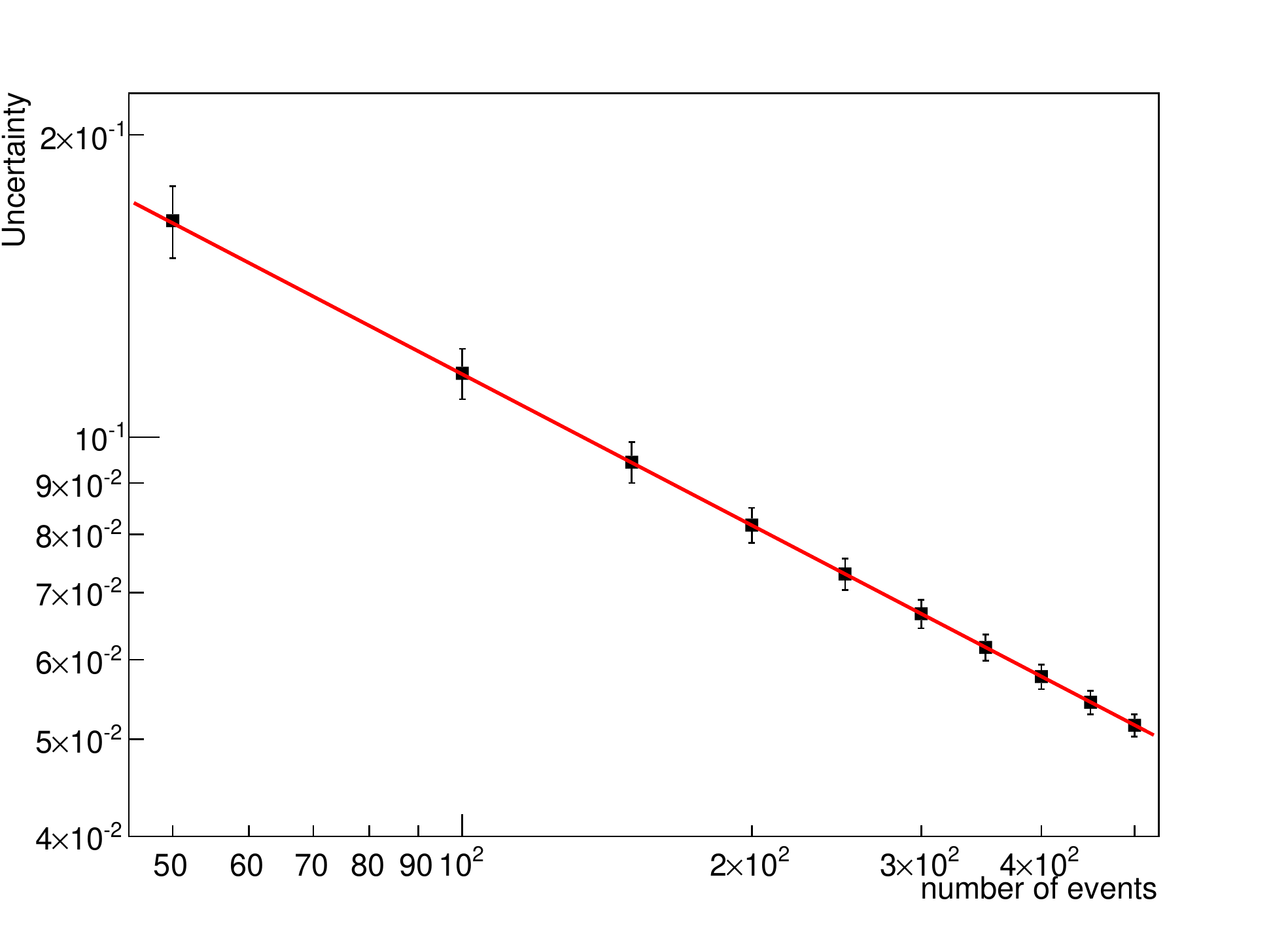}
            \caption{Uncertainty $\est{\sigma}_5$ of the angular
              observable $S_5$ extracted from $2\cdot 10^5$ studies of
              simulated events for the decay $B\to
              K^*\ell^+\ell^-$. We show the uncertainty as a function
              of the number of simulated events $N$.  The red curve
              represents a fit to the function given in
              \refeq{unc-on-mean}. The error bars correspond to the
              $68\%$ spread of measured uncertainties in the toys. The
              plot for $\est{\sigma}_7$ is visually indistinguishable
              from the one shown here.}
        \label{fig:errors}
\end{figure}

\begin{figure}[t]
        \centering
            \includegraphics[width=0.45\textwidth]{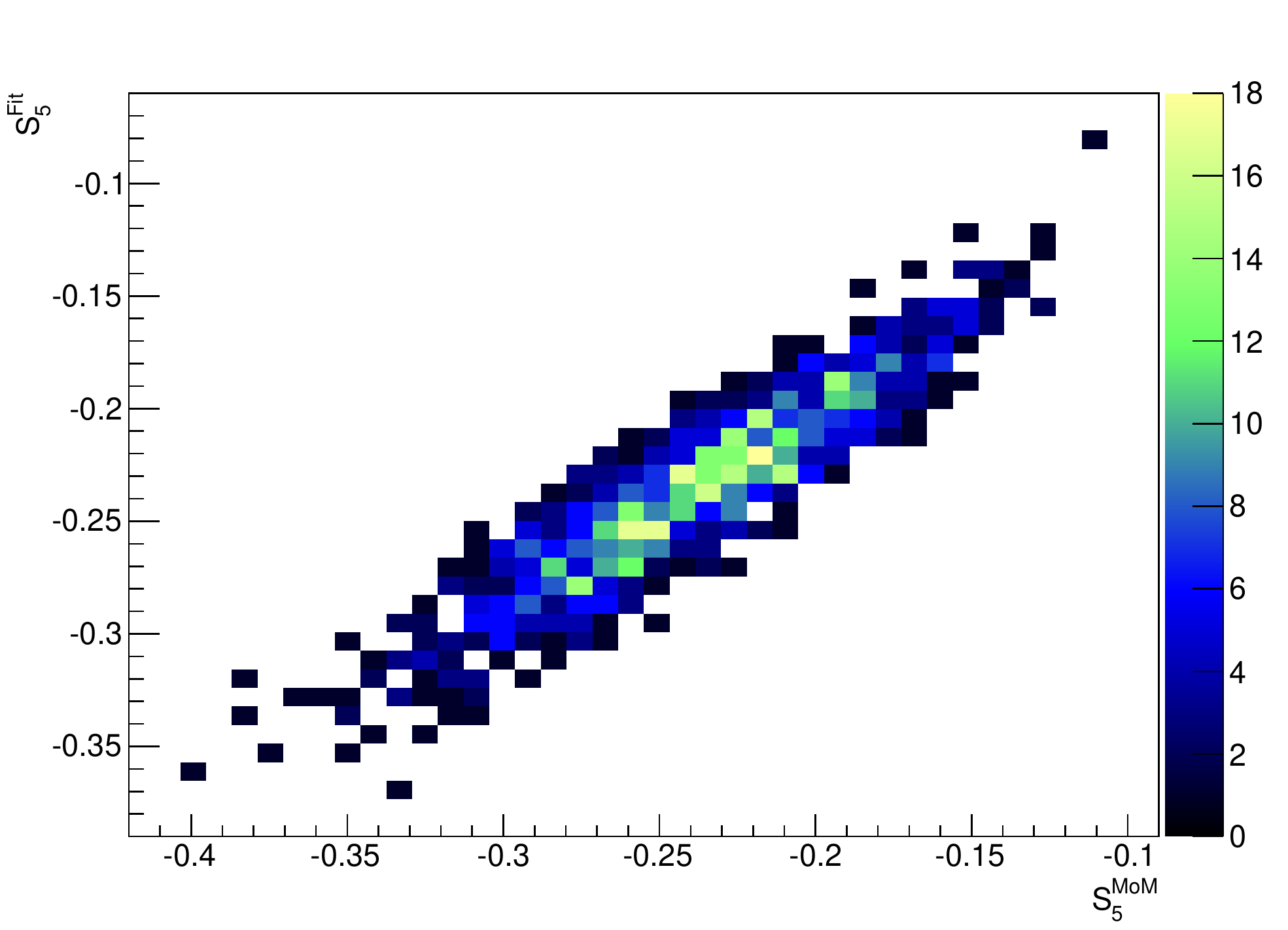}
            \includegraphics[width=0.45\textwidth]{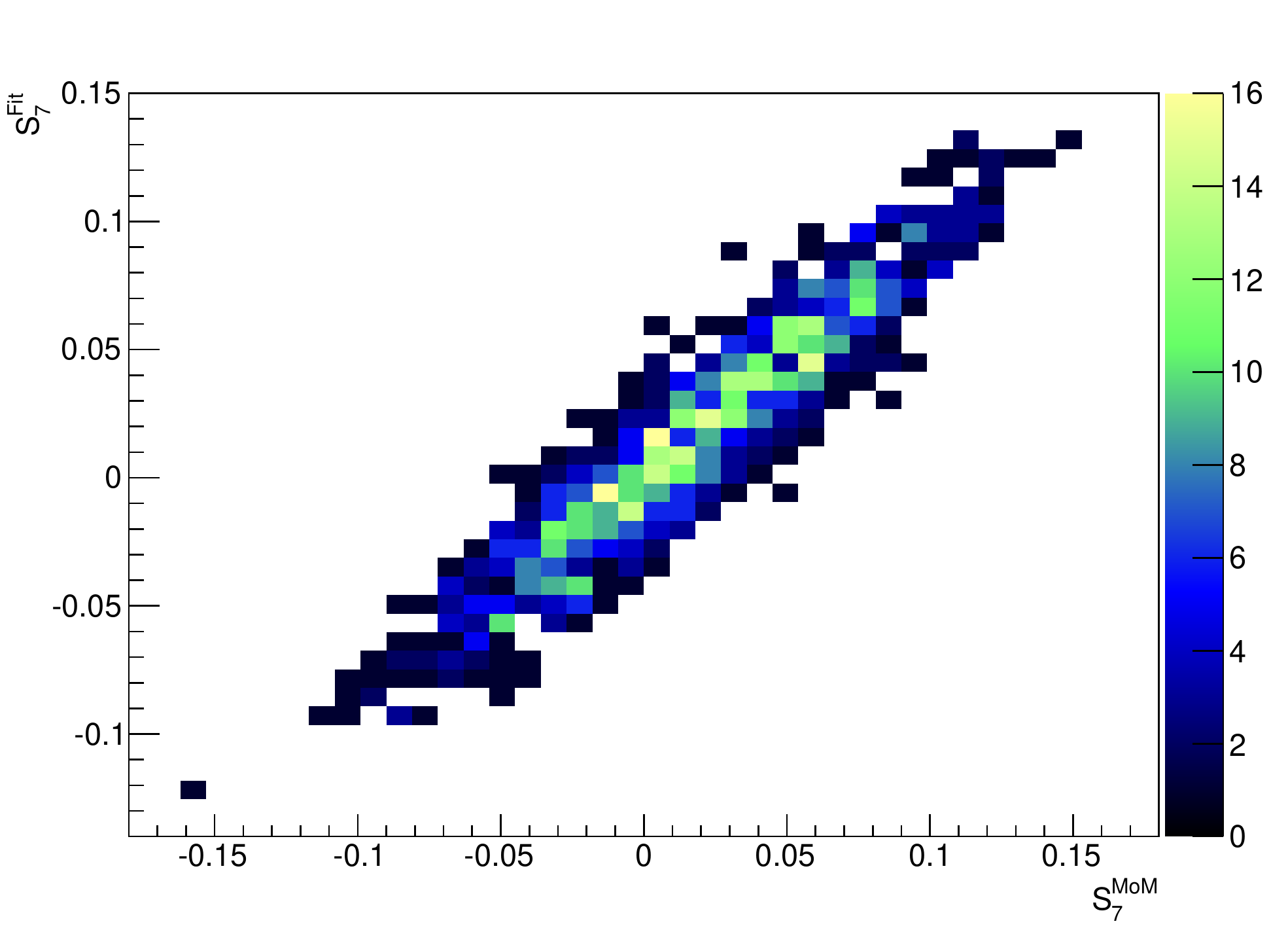}
            \caption{The joint distribution of the two estimators
                $\est{S}_i^\text{MoM}$ and $\est{S}_i^\text{Fit}$ that arise
                from the method of moments and a maximum-likelihood fit,
                respectively. We show the estimators for both of our benchmark
                observables $S_5$ (upper) and $S_7$ (lower).
            }
        \label{fig:correlation}
\end{figure}

Our findings can be summarized as follows:
\begin{itemize}
    \item In all studied cases we observed not a single bias in the distribution of the $\mathrm{pull}$
        of any observables $S_i$,
        \begin{equation}
            \mathrm{pull_i} \equiv \frac{\est{S}_i - S_i}{\est{\sigma}_{i}}\,.
        \end{equation}
        Here $S_i$ refers to the true (input) value for the angular observables, $\est{S}_i$ refers to the
        mean of the pseudo measurement via the MC estimate, and $\est{\sigma}_{i} \equiv \est{\Sigma_{ii}}^{1/2}$
        refers to an estimator of the standard deviation of the pseudo measurement. All pull distributions
        obtained in our studies can be successfully
        fitted to a Gaussian distribution. Out of the large number  of studied distributions, we only show the pull distributions for the observables
        $S_5 \simeq 27\%$ and $S_7 \simeq 2\%$ obtained from SM-like $B\to K^*\ell^+\ell^-$
        decays as representative examples. We generated $2\cdot 10^5$ toy studies
        with $200$ events per study in \reffig{pulls}.

    \item We study the MC estimate for the absolute uncertainty $\est{\sigma_i(N)}$ with respect to the angular observable $S_i$
        as a function of the number of simulated events $N$. As expected for a multivariate Gaussian
        distribution, we find that the absolute uncertainty is well fitted by
        \begin{equation}
            \label{eq:unc-on-mean}
            \est{\sigma_i(N)} = \frac{\sigma_i}{\sqrt{N}}
        \end{equation}
        with $\sigma_i(1) = \order{1}$, regardless of the absolute size of $S_k$. The latter can best be shown
        for the example of uncertainties of two observables. Taking again $S_5$ ($\simeq 27\%$) and $S_7$ ($\simeq 2\%$)
        for SM-like $B\to K^*\ell^+\ell^-$ decays, we show the absolute uncertainty in \reffig{errors}.
        We find that the the method of moments yields uncertainties on $\vec{S}$ that are roughly
        $10\%$ -- $30\%$ larger than those obtained from maximum-likelihood fits and for the same
        number of events. However, we wish to note that said fits only produce a limited
        subset of the angular observables, and the statistical error of the fit is expected to increase
        with the number of fit parameters until their errors saturate the statistical errors of the method-of-moments
        estimators \cite[sec. 8]{Cowan:1998ji}.

    \item We also compare the results as obtained by the method of moments with
        results obtained by a conventional likelihood fit. In particular, we
        study the correlation between the method-of-moment estimators and the
        maximum-likelihood-fit estimators. We run $10^3$ toy analyses, with
        $200$ simulated events per analysis. We show the joint distribution of
        the two estimators in \reffig{correlation}. The two estimators are
        highly correlated. The distribution of the difference of the estimators
        exhibits now bias, which is due to the large number of simulated
        events.  Still, we find that statistical uncertainty on the difference
        of the two estimators is sizeable and can easily become half as large
        as the statistical uncertainty of either estimator.
\end{itemize}

We emphasize that the above results have been obtained for a flat acceptance
function.  We also study the behavior of the unfolded angular observables.
For simplicity we limit our study to the decay $B\to K\ell^+\ell^-$, with its
three angular observables $S_0$ to $S_2$. We express the acceptance function
in terms of Legendre polynomials $p_k(x)$,
\begin{equation}
    \eps(\cos\theta) = \frac{7}{15} p_0(\cos\theta) - \frac{4}{15} p_2(\cos\theta)\,.
\end{equation}
This acceptance function approximates the one used in a recent study of the
angular observables in $B\to K\ell^+\ell^-$ decays \cite{Aaij:2014tfa}.
Focusing on effects of the unfolding process itself, we use the analytical
expression for the unfolding matrix, which we compute from the raw moments as
defined in \refeq{analytic-raw-moments}. Simulting $4000$ toy analyses with
up to $300$ simulated events each, we find that the previous bullet points
still hold; i.e., we do not find any bias, and the distribution is well
described by a multivariate Gaussian distribution. The latter only holds as
long as the number of events per experiment exceeds $\sim 30$.\\

All of our toy studies, as summarized above, show consistently that the
joint distribution of the angular observables converges rapidly towards a
multivariate Gaussian distribution. We therefore propose to publish the results
in the form of the physical components of $\vecest{S}$ and $\est{\Sigma}$.

\section{Summary}

We have carried out a combined analytical and numerical study of the method of
moments; a method for the extraction of angular observables from the angular
distribution of a general multi-body process.  We have studied the performance
of the method of moments using pseudo data derived from the SM predictions for
one uniangular decay ($B\to K\ell^+\ell^-$) and one triangular decay ($B\to
K^*\ell^+\ell^-$).  From this, we find rapid convergence of the joint
likelihood of the angular observables towards a multivariate Gaussian.  We draw
the conclusion that this method exhibits several benefits in the determination
of angular observables when compared with a maximum-likelihood fit.\\

First, we find no bias in the determinations of the angular
observables even for a small number of events.  However, due to fewer
model assumptions, the uncertainty on the mean values increases by
roughly $10\%$--$30\%$ compared
to likelihood fits.\\
Second, in the absence of detector effects, the method of moments does
not rely on model assumptions for the partial-wave composition of the
PDF. This is explicitly shown for the case of higher
partial waves in multibody final states.\\
Third, we develop a systematic method for the determination of
detector effects that lead to dilution and mixing of the angular
observables. We present an algorithm to calculate the necessary
unfolding matrix, which is computationally feasible only when using
the method
of moments. The algorithm also accounts for higher partial waves.\\
Fourth, the joint distribution of the angular observables resulting
from the method of moments is well approximated by a multivariate
Gaussian distribution even for small number of events $N \sim 30$ both
for the ideal uniform acceptance and a realistic example.  This
facilitates the precise transfer of correlation information
to subsequent theoretical analyses.\\
Last but not least, the resulting distribution arises without the need
for additional model constraints. Thus more observables can be
inferred from the same data than in a likelihood fit. In addition,
the results from the method of moments can be more easily averaged or combined; e.g., in global fits.\\

In conclusion, we argue that the method of moments is a competitive alternative to
maximum-likelihood fits if angular distributions are involved. We wish to raise
the interesting prospect of extending this method to applications that feature
PDFs composed from non-angular orthogonal bases.

\acknowledgments

We thank Ulrik Egede for insightful discussions and helpful comments on the
manuscript. We are grateful to Robert Fleischer, Gudrun Hiller, Martin Jung and
Konstantinos Petridis for helpful discussions, as well as Claus Grupen for
helpful advice in searching the literature.
N.S. acknowledges the support of the Swiss National Science Foundation,
PP00P2-144674.
The work of D.v.D has been supported by the Bundesministerium f\"ur Bildung und
Forschung (BMBF).

\appendix

\section{Application to $\bar{B}\to\bar{K}\ell^+\ell^-$}
\label{app:btokll}

The PDF for the decay $\bar{B}\to\bar{K}\ell^+\ell^-$ has been calculated for the most
complete basis of dimension-six $b\to s \ell^+\ell^-$ operators. It reads \cite{Bobeth:2007dw,Bobeth:2012vn}
\begin{equation}
\begin{aligned}
    P(q^2, \cos\theta_1)
        & = \frac{1}{\rmdx{\Gamma}/\rmdx{q^2}} \frac{\rmdx{^2\Gamma}}{\rmdx{q^2} \rmdx{\cos\theta_1}}\\
        & = \frac{a(q^2)}{\rmdx{\Gamma}/\rmdx{q^2}} + \frac{b(q^2)}{\rmdx{\Gamma}/\rmdx{q^2}} \cos\theta_1\\
        & \quad + \frac{c(q^2)}{\rmdx{\Gamma}/\rmdx{q^2}} \cos^2\theta_1\\
        & \equiv \sum_i S_i p_i(\cos\theta_1)\,,
\end{aligned}
\end{equation}
with the conventional observables $a(q^2)$ through $c(q^2)$, and $\rmdx{\Gamma}/\rmdx{q^2} = 2a + 2/3 c$. We conveniently use the Legendre polynomials
$p_i(x)$, $i=0,1,2$,
\begin{equation}
\begin{aligned}
    p_0(x) & = 1\,, &
    p_1(x) & = x\,, &
    p_2(x) & = \frac{1}{2} (3x^2 - 1)\,,
\end{aligned}
\end{equation}
as our basis of angular functions. Our basis of angular observables then
translates to the conventional basis as
\begin{equation}
\begin{aligned}
    S_0 & = \frac{1}{2}\,, &
    S_1 & = \frac{b}{\Gamma}\,, &
    S_2 & = \frac{2c}{3\Gamma}\,.
\end{aligned}
\end{equation}
In this case, the dual basis is simply given by $\tilde{p}_i(x) = (2 i + 1)/2 p_i(x)$
such that
\begin{equation}
    \int_0^\pi \rmdx{\theta_1} \dual{p}_i(\cos\theta_1) P(q^2, \cos\theta_1)\sin\theta_1 = S_i(q^2)\,.
\end{equation}

\section{Application to $\Lambda_b\to \Lambda(\to N \pi)\ell^+\ell^-$}
\label{app:lambdabtolambdall}

The PDF for the decay --- in the presence of Standard-Model operators and their chirality-flipped counter parts --- reads \cite{Boer:2014kda}
\begin{equation}
\begin{aligned}
    P(q^2, \cos\theta_1, \cos\theta_2, \theta_3)
        & = \frac{(\rmdx{\Gamma} /\, \rmdx{q^2})^{-1} \rmdx{^4\Gamma}}{\rmdx{q^2} \rmdx{\cos\theta_1} \rmdx{\cos\theta_2} \rmdx{\theta_3}}\\
        & = \sum_i S_i f_i(\cos\theta_1, \cos\theta_2, \theta_3)\,,
\end{aligned}
\end{equation}
where $q^2$ denotes the dilepton mass squared, $\theta_1 \equiv \theta_\ell$ and $\theta_2 \equiv \theta_\Lambda$ denote the
helicity angles in the dilepton and
$N\pi$ systems, respectively, and $\theta_3 = \phi$ denotes the azimuthal angle.
The index $i$ should be interpreted as a multi-index, $i \equiv (l_1, l_2, m)$,
where $0 \leq l_1 \leq 2$ and $0 \leq l_2 \leq 1$ denote the total angular momentum in the
dilepton and the $N\pi$ system, respectively, and $-1 \leq m \leq 1$
is the third component of either of the angular momenta.\\

Our choice of an orthonormal basis reads
\begin{multline}
\label{eq:lambdab:bases1}
    f_{l_1, l_2, m}(\cos\theta_1, \cos\theta_2, \theta_3)\\
    = \sqrt{\frac{(l_1 - |m|)!\,(l_2 - |m|)!}{(l_1 + |m|)!\,(l_2 + |m|)!}}\\
    \quad \times p_{l_1}^{|m|}(\cos\theta_1) p_{l_2}^{|m|}(\cos\theta_2)\\
    \quad \times \begin{cases}
            \cos(|m|\theta_3) & m > 0\\
            \sin(|m|\theta_3) & m < 0\\
            1                 & m = 0
        \end{cases}\,,
\end{multline}
and its dual is
\begin{multline}
\label{eq:lambdab:bases2}
    \tilde{f}_{l_1, l_2, m}(\cos\theta_1, \cos\theta_2, \theta_3)\\
    = \frac{(2l_1 + 1)(2l_2 + 1)}{8\pi}\sqrt{\frac{(l_1 - m)!\,(l_2 - m)!}{(l_1 + m)!\,(l_2 + m)!}}\\
    \quad\times p_{l_1}^m(\cos\theta_1) p_{l_2}^m(\cos\theta_2)\\
    \quad \times \begin{cases}
            2 \cos(|m|\theta_3) & m > 0\\
            2 \sin(|m|\theta_3) & m < 0\\
            1                   & m = 0
        \end{cases}\,.\\
\end{multline}

The correspondence between out choice of angular observables in the angular momentum basis, and the angular
observables as defined in reference \cite{Boer:2014kda} reads
\begin{widetext}
\begin{equation}
\begin{aligned}
    8\pi S_{0, 0,  0} & = 1\,,                                                             &
    8\pi S_{0, 1,  0} & = \frac{K_{2cc} + 2 K_{2ss}}{\rmdx\Gamma/\rmdx{q^2}}\,,
\end{aligned}
\end{equation}
and
\begin{equation}
\begin{aligned}
                      &                                                                    &
    8\pi S_{1, 1, -1} & = \frac{6 K_{4s}}{\rmdx\Gamma/\rmdx{q^2}}\,,                       \\
    8\pi S_{1, 0,  0} & = \frac{3 K_{1c}}{\rmdx\Gamma/\rmdx{q^2}}\,,                       &
    8\pi S_{1, 1,  0} & = \frac{3 K_{2c}}{\rmdx\Gamma/\rmdx{q^2}}\,,                       \\
                      &                                                                    &
    8\pi S_{1, 1, +1} & = \frac{6 K_{3s}}{\rmdx\Gamma/\rmdx{q^2}}\,,
\end{aligned}
\end{equation}
and
\begin{equation}
\begin{aligned}
                      &                                                                    &
    8\pi S_{2, 1, -1} & = \frac{2 \sqrt{3} K_{4sc}}{\rmdx\Gamma/\rmdx{q^2}}\,,             \\
    8\pi S_{2, 0,  0} & = \frac{2(K_{1cc} - K_{1ss})}{\rmdx\Gamma/\rmdx{q^2}}\,,           &
    8\pi S_{2, 1,  0} & = \frac{2(K_{2cc} - K_{2ss})}{\rmdx\Gamma/\rmdx{q^2}}\,,           \\
                      &                                                                    &
    8\pi S_{2, 1, +1} & = \frac{2 \sqrt{3} K_{3sc}}{\rmdx\Gamma/\rmdx{q^2}}\,,
\end{aligned}
\end{equation}
where the decay width is
\begin{equation}
    \frac{\rmdx{\Gamma}} {\rmdx{q^2}} = 2 K_{1ss} + K_{1cc}\,.
\end{equation}

The dual basis is chosen such that
\begin{equation}
    \int_{-1}^{+1} \rmdx{\cos \theta_1} \int_{-1}^{+1} \rmdx{\cos \theta_2} \int_0^{2\pi} \rmdx{\theta_3} P(q^2, \cos \theta_1, \cos \theta_2, \theta_3) \dual{f}_i(\cos \theta_1, \cos \theta_2, \theta_3) = S_i(q^2)\,.
\end{equation}
\end{widetext}

For the purpose of unfolding acceptance effects as laid down in \refsec{systematics:acceptance}, it is instrumental
to know that $f_{0,0,0} \equiv 1$, and that $\max_{\cos\theta_1,\cos\theta_2,\phi} |f_{l_1, l_2, m}| < 1$.\\

The recipe's generating PDFs are therefore $P(x|\lbrace S_i^{(j)}\rbrace)$, with
\begin{equation}
    \label{eq:recipe:lambdabtolambdall}
        S_i^{(j)} = \frac{1}{8\pi}
        \begin{cases}
            \delta_{i,(0,0,0)}                  & j = (0, 0, 0)\\
            \delta_{i,(0,0,0)}  + \delta_{i,j}  & j \neq (0, 0, 0)
        \end{cases}\,,
\end{equation}
and where $j$ is now also a multi-index representing $j \equiv (\tilde l_1, \tilde l_2, \tilde m)$.

\section{Application to $\bar{B}\to\bar{K}\pi\ell^+\ell^-$}
\label{app:btokstarll}

The PDF for the decay $\bar{B}\to\bar{K}\pi\ell^+\ell^-$ --- up to and including P-wave contributions --- has been calculated
for the most general basis of dimension-six $b\to s$ operators. It reads, expressed in terms of the angular observables $\lbrace J_i\rbrace$ \cite{Blake:2012mb,Bobeth:2012vn}
\begin{equation}
\begin{aligned}
    P(q^2, \cos\theta_1, \cos\theta_2, \theta_3)
    & = \frac{(\rmdx{\Gamma}/\rmdx{q^2})^{-1}\rmdx{^4\Gamma}}{\rmdx{q^2} \rmdx{\cos\theta_1} \rmdx{\cos\theta_2} \rmdx{\theta_3}}\\
    & = \sum_i S_i(q^2) f_i(\cos\theta_1, \cos\theta_2, \theta_3)\,,
\end{aligned}
\end{equation}
where $\theta_1 \equiv \theta_\ell$ is the dilepton helicity angle; $\theta_2 \equiv \theta_{K}$ is the $\bar{K}\pi$ helicity angle; $\theta_3 \equiv \phi$ is the azimuthal angle;
and $q^2$ is the square of the dilepton mass. The $q^2$-differential decay width reads
\begin{equation}
    \frac{\rmdx{\Gamma}}{\rmdx{q^2}} = \frac{\big(3 J_{1c} - J_{2c}\big) + 2\big(3J_{1s} - J_{2s}\big)}{3}\,.
\end{equation}
It is convenient to define the basis of angular functions and its dual in terms of
associated Legendre polynomials $p_l^m(x)$. The index $i$ should thus be interpreted as a multi-index
, $i \equiv (l_1, l_2, m)$, where $0 \leq l_1 \leq 2$ and $0 \leq l_2 \leq 2$ denote the total angular
momentum in the dilepton and the $K\pi$ system, respectively, and $-2 \leq m \leq 2$
is the third component of either of the angular momenta.
We use the same bases of angular functions as given in \refeq{lambdab:bases1} and \refeq{lambdab:bases2}
for the decay $\Lambda_b\to \Lambda\ell^+\ell^-$.
\begin{widetext}
In that case, the angular observables correspond to the usual choice of observables via
\begin{equation}
\begin{aligned}
    8\pi S_{0, 0,  0} & = 1\,,                                                                             &
    8\pi S_{0, 1,  0} & = \frac{3 J_{1i} - J_{2i}}{\rmdx{\Gamma}/\rmdx{q^2}}\,,                            &
    8\pi S_{0, 2,  0} & = \frac{6 (J_{1c} - J_{1s}) - 2(J_{2c} - J_{2s})}{3\rmdx{\Gamma}/\rmdx{q^2}}\,,
\end{aligned}
\end{equation}
and
\begin{equation}
\begin{aligned}
                      &                                                                                    &
    8\pi S_{1, 1, -1} & = \frac{6 J_{7i}}{\rmdx{\Gamma}/\rmdx{q^2}}\,,                                     &
    8\pi S_{1, 2, -1} & = \frac{4 \sqrt{3} J_{7}}{\rmdx{\Gamma}/\rmdx{q^2}}\,,                             \\
    8\pi S_{1, 0,  0} & = \frac{J_{6c} + 2 J_{6s}}{\rmdx{\Gamma}/\rmdx{q^2}}\,,                            &
    8\pi S_{1, 1,  0} & = 0\,,                                                                             &
    8\pi S_{1, 2,  0} & = \frac{2(J_{6c} - J_{6s})}{\rmdx{\Gamma}/\rmdx{q^2}}\,,                           \\
                      &                                                                                    &
    8\pi S_{1, 1, +1} & = \frac{6 J_{5i}}{\rmdx{\Gamma}/\rmdx{q^2}}\,,                                     &
    8\pi S_{1, 2, +1} & = \frac{4 \sqrt{3} J_{5}}{\rmdx{\Gamma}/\rmdx{q^2}}\,,
\end{aligned}
\end{equation}
and
\begin{equation}
\begin{aligned}
                      &                                                                                    &
                      &                                                                                    &
    8\pi S_{2, 2, -2} & = \frac{8 J_9}{\rmdx{\Gamma}/\rmdx{q^2}}\,,                                        \\
                      &                                                                                    &
    8\pi S_{2, 1, -1} & = \frac{4 \sqrt{3} J_{8i}}{\rmdx{\Gamma}/\rmdx{q^2}}\,,                            &
    8\pi S_{2, 2, -1} & = \frac{8 J_8}{\rmdx{\Gamma}/\rmdx{q^2}}\,,                                        \\
    8\pi S_{2, 0,  0} & = \frac{4 (J_{2c} + 2 J_{2s})}{3\, \rmdx{\Gamma}/\rmdx{q^2}}\,,                    &
    8\pi S_{2, 1,  0} & = \frac{4 J_{2i}}{\rmdx{\Gamma}/\rmdx{q^2}}\,,                                     &
    8\pi S_{2, 2,  0} & = \frac{8 (J_{2c} - J_{2s})}{3\, \rmdx{\Gamma}/\rmdx{q^2}}\,,                      \\
                      &                                                                                    &
    8\pi S_{2, 1, +1} & = \frac{4\sqrt{3} J_{4i}}{\rmdx{\Gamma}/\rmdx{q^2}}\,,                             &
    8\pi S_{2, 2, +1} & = \frac{8 J_{4}}{\rmdx{\Gamma}/\rmdx{q^2}}\,,                                      \\
                      &                                                                                    &
                      &                                                                                    &
    8\pi S_{2, 2, +2} & = \frac{8 J_3}{\rmdx{\Gamma}/\rmdx{q^2}}\,.                                        \\
\end{aligned}
\end{equation}
\end{widetext}

As before, for the purpose of unfolding acceptance effects as laid down in \refsec{systematics:acceptance}, it is instrumental
to know that $f_{0,0,0} \equiv 1$, and that $\max_{\cos\theta_1,\cos\theta_2,\phi} |f_{l_1, l_2, m}| \leq 1$.
The recipe's generating PDFs are therefore $P(x|\lbrace S_i^{(j)}\rbrace)$ with
\begin{equation}
    \label{eq:recipe:btokstarll}
    S_i^{(j)} = \frac{1}{8\pi}
        \begin{cases}
            \delta_{i,(0,0,0)}                  & j = (0, 0, 0)\\
            \delta_{i,(0,0,0)}  + \delta_{i,j}  & j \neq (0, 0, 0)
        \end{cases}
\end{equation}
where also $j$ is now a multi-index representing $j \sim (l_1, l_2, m)$.

To conclude this section, we remind that the angular momentum $l_2$, associated with the angle $\theta_2$, is not bounded from above.
This has to be considered if partial waves beyond the P-wave are included in the analysis, see e.g. \cite{Das:2014sra}.
However, our choice of basis is well suited to these applications with $0 \leq l < \infty$. Note, that the
physical range of $m$ is not affected by higher partial waves.\\

\section{On the Partial-Wave Expansion of Angular Observables}
\label{app:partial-waves}

For convenience, we use this appendix to collect there necessary
formulae needed in the partial-wave expansion.
Let us assume that the angular decomposition has been achieved for
some PDF $P$ for all angles except for one angle $\theta$. We now
focus on just one of the resulting observables and denote it by
$S(\theta)$. The dependence on the non-angular variables will be
ignored in the following. Suppose $S$
has an expansion in terms of partial waves $l_1, l_2 = 0,1,2,\dots
\hat{=}$ S,P,D$,\dots$ of the underlying amplitudes $A_1$ and $A_2$,
\begin{widetext}
\begin{equation}
    \label{eq:def-partial-wave-observable}
    S(\theta) \equiv F\left[A_1(\theta) A_2^*(\theta)\right] \equiv F\left[\left(\sum_{l_1=0}^\infty A_1^{(l_1)} p_{l_1}^{(m_1)}(\cos\theta)\right) \left(\sum_{l_2=0}^\infty A_2^{*(l_2)} p_{l_2}^{(m_2)}(\cos\theta)\right)\right]\,,
\end{equation}
\end{widetext}
where $F \in \{\text{Re},\text{Im}\}$ denotes taking either the real
or the imaginary part. Here $p_{l}^{(m)}$ denotes an \emph{associated Legendre} polynomial and $m_i$ is the third component of the angular momentum of the amplitude $A_i$, and we impose $m_1 \geq m_2$.\\

From the orthogonality of the Legendre polynomials, one immediately finds
\begin{equation}
    \label{eq:partial-waves:bound}
    \int \rmdx{\cos\theta}  \, |A_i(\theta)|^2 = \sum_{l_i = 0}^\infty |A_i^{(l_i)}|^2 n_{l_i, m_i} < S_\text{incl}\,,
\end{equation}
where $S_\text{incl}$ is the corresponding observable in the associated inclusive decays,
and where we introduce $n_{l,m}$ via the scalar product of two associated Legendre polynomials,
\begin{equation}
\label{eq:legendre-scalar-product}
\begin{aligned}
    n_{l, m} \delta_{l, l'}
    & \equiv \int_{-1}^1 \rmdx{\cos\theta} p_{l}^{(m)}(\cos\theta) p_{l'}^{(m)} (\cos\theta)\\
    & = \frac{2}{(2 l + 1)} \frac{(l + m)!}{(l - m)!} \delta_{l, l'}\,.
\end{aligned}
\end{equation}
The positivity of the amplitudes in \refeq{partial-waves:bound} implies that we can estimate
the error introduced by cutting off the expansion at some arbitrary angular momentum $L$. One
obtains
\begin{equation}
    |A_i^{(l' > L)}|^2 < S_\text{incl} - \sum_{l=0}^L |A_i^{(l)}|^2\,.
\end{equation}

Will will show in the following that such a cutoff is compatible with defining a basis of angular
observables as coefficients of Legendre polynomials in $\cos\theta$. Since this expansion implies
a well defined total angular momentum for each observable, one ensures that the observables can in
fact be disentangled experimentally.\\

We decompose $S$ in terms of the associated Legendre polynomials $p_{j}^{(m)}(\cos\theta)$,
with total angular momentum $j$ and its third component $m=m_1 + m_2$.
\begin{equation}
    S(\theta) = \sum_j S_{j,m} p_{j}^{(m)}(\cos\theta)\,.
\end{equation}
This parametrization has two merits. First, we can immediately project out the angular observables $S_{j,m}$ by means of \refeq{legendre-scalar-product}:
\begin{equation}
    S_{j,m} = \frac{1}{n_{j,m}} \int_{-1}^{+1} \rmdx{\cos\theta} S(\theta) p_{j}^{(m)}(\cos\theta)\,.
\end{equation}
(Here and in the next step we may exchange the integral and the
series because each element of the series is a product of
polynomials on the compact support [-1,1], and thus each integral
is absolutely convergent).
Second, we can immediately express $S_{j,m}$ in terms of
the partial-wave amplitudes,
\begin{widetext}
\begin{equation}
    \label{eq:partial-wave-observable-infinite}
    \begin{aligned}
        S_{j,m}
            & = \frac{1}{n_{j,m}} \int_{-1}^{+1} \rmdx{\cos\theta} p_{j}^{(m_1 + m_2)}(\cos\theta) \sum_{l_1,l_2=0}^\infty F\left[A_{1}^{(l_1)} p_{l_1}^{(m_1)}(\cos\theta) A_{2}^{*(l_2)}p_{l_2}^{(m_2)}(\cos\theta)\right]\\
            & = \sum_{l_1,l_2=0}^\infty F\left[A_{1}^{(l_1)} A_{2}^{*(l_2)}\right] \frac{T_{l_1,l_2,j}^{(m_1,m_2)}}{n_{j,m}}\,,\qquad\text{with }m = m_1 + m_2\,.
    \end{aligned}
\end{equation}
In the last step, we use Gaunt's formula \cite{Gaunt:1929} to integrate a triple product of associated Legendre polynomials,
\begin{equation}
\begin{aligned}
    T_{l_1,l_2,j}^{(m_1,m_2)}
        & = \int_{-1}^{+1} \rmdx{\cos\theta} p_{j}^{(m_1 + m_2)}(\cos\theta) p_{l_1}^{(m_1)}(\cos\theta) p_{l_2}^{(m_2)}(\cos\theta)\\
        & = (-1)^{s - l_1 - m_2} \frac{2 (l_1 + m_1)! (l_2 + m_2)! (2s - 2 l_2)! s!}{(l_1 - m_1)! (s - j)! (s - l_1)! (s - l_2)! (2s + 1)!}\\
        & \times \sum_{t=p}^q (-1)^t \frac{(j + m + t)!(l_1 + l_2 - m - t)!}{t! (j - m - t)! (l_1 - l_2 + m + t)! (l_2 - m_2 - t)!}\,,
\end{aligned}
\end{equation}
\end{widetext}
where
\begin{equation}
\begin{aligned}
    m & = m_1 + m_2\,, &
    m_1 & \geq m_2\,,  \\
    j, l_1, l_2 & \geq 0\,, &
    m, m_1, m_2 & \geq 0\,,
\end{aligned}
\end{equation}
and
\begin{equation}
\begin{aligned}
    s & \equiv \frac{j + l_1 + l_2}{2}\,, \\
    p & \equiv \max(0, l_2 - l_1 - m)\,, \\
    q & \equiv \min(l_1 + l_2 - m, j - m, l_2 - m_2)\,.
\end{aligned}
\end{equation}
The necessary conditions for $T \neq 0$ are
\begin{equation}
    \label{eq:angular-momentum-addition}
    s \in \mathbb{N}\qquad \wedge \qquad l_1 - l_2 \leq l \leq l_1 + l_2,.
\end{equation}
The latter condition is well known from the addition rules of angular momenta. Note, however, that
the sum in \refeq{def-partial-wave-observable} goes to infinitely high angular momenta $l_1$ and $l_2$. As a consequence
of this and of \refeq{angular-momentum-addition}, the angular observables $S_{j,m}$
consist of sums with infinitely many terms. It is then up to theoretical analyses to
estimate or calculate the impact of the neglected partial waves, e.g. as outlined above.

\bibliography{references}

\end{document}